\documentclass{article}

\usepackage{amssymb,amsmath}
\usepackage{exscale,graphicx}

\newcommand{\bra}[1]{\langle #1 |}
\newcommand{\ket}[1]{| #1 \rangle}

\begin{document}

\title{\bf
   Continuous history variable for
   programmable quantum processors 
}

\author{
  Alexander Yu.\ Vlasov
}

\maketitle

\begin{abstract}
In this brief note is discussed application of continuous quantum
history (``trash'') variable for simplification of scheme of 
programmable quantum processor. Similar scheme may be tested also in 
other models of the theory of quantum algorithms and complexity, 
because provides modification of a standard operation: {\em quantum function
evaluation}. 
\end{abstract}



\section{Preliminaries}

It was discussed in \cite{NC97}, that programmable quantum computer
may be universal only in approximate sense. On the other hand, such
computer may approximate any operation with arbitrary precision, if
to repeat an elementary step sufficient number of times (``timing'') 
\cite{AV01,AV03,AV07}. 
In fact, such kind of universality is rather standard since first papers 
about quantum computational networks \cite{QCN}. Experimental realizations
of programmable quantum computers also may use similar idea \cite{PQC}.

Really, programmable quantum processor sometimes could be compared with 
usual classical computer, controlling sequence of applications of
quantum gates. There is well known formal method of revision of a
scheme of a classical computation into a model, compatible with
quantum laws. It is possible first to use some reversible design 
of Turing machine \cite{BenRev}. The quantum mechanical model of 
such device is quite straightforward \cite{BenEr}.

In quantum circuit model similar approach sometimes is denoted
as ``quantum function evaluation'' \cite{QAlg}, but it is 
rather classical idea with using instead of irreversible function 
$f(x)$ the reversible one on the {\em pair of arguments}, like
$\tilde f : (x,y) \mapsto (x,f(x) \circleddash y)$, where for
different domains of $x$ operator $\circleddash$ may be 
subtraction (in modular or usual arithmetics) or
{\em bitwise exclusive or} (for binary case). The $\tilde f$ 
is reversible ($\tilde f$ is involution, {\em viz} 
$\tilde f^{(-1)} = \tilde f$) and
\begin{equation}
\tilde f : (x,0) \mapsto (x,f(x)) 
\label{revf}
\end{equation}

\begin{figure}[htb]
\begin{center}
\includegraphics[scale=0.75]{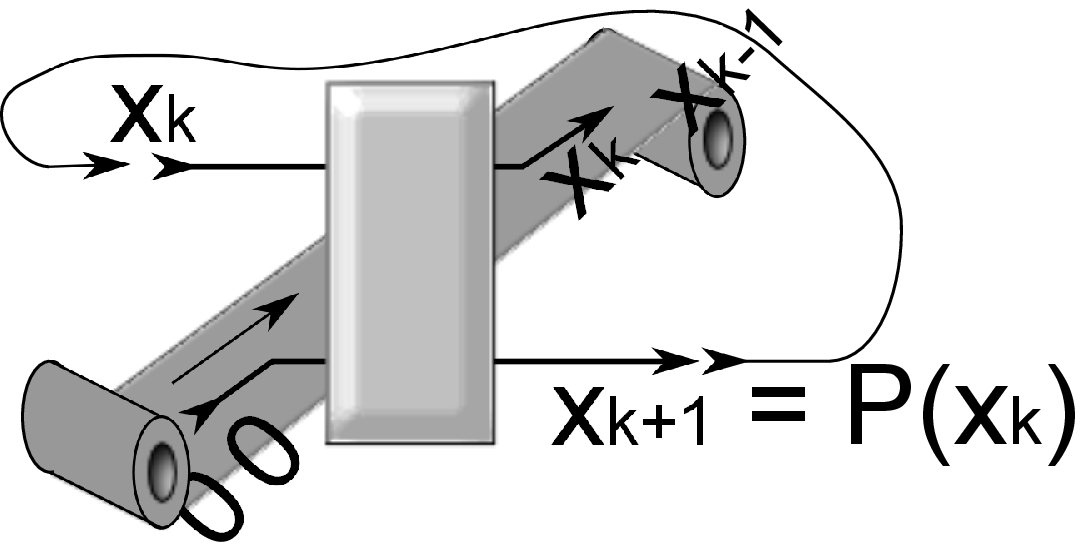} 
\end{center}
\caption{Scheme of tapes}\label{Fig:tapes}
\end{figure}

But such a transition produces certain problem with ``timing''. 
An initial irreversible computer might be considered as a sequence 
$M \to f(M) \to f(f(M)) \cdots $, there
$M$ is state of whole memory (containing both data and program) 
and $f$ is a fixed ``function'' corresponding to a 
circuits design.

For reversible function described above in Eq.~(\ref{revf}) 
it is necessary on each step to provide new fresh input with zeros 
and to withdraw the redundant copy of $x$ from the output
(see Figure~\ref{Fig:tapes}). 
It is analogue of two additional tapes in Turing machine design 
\cite{BenRev}. For circuit model it corresponds to the growth of 
additional memory resources linearly with maximal number of steps, 
necessary to perform required task.

\section{Brief description}

To resolve this problem here is suggested a model 
of encoding both tapes into single {\em continuous quantum variable} 
\cite{contvar}.
Let us consider discrete quantum variable $\ket{b}$ ({\em e.g.}, a 
qubit may be considered {\em w.l.o.g.})
and continuous one $\ket{x}$. 

It may be compared with classical
example, then two semi-infinite tapes (zeros and history) are 
encoded into single real number represented in binary notation as
$x~=~\ldots z_{n+2}z_{n+1}.b_n b_{n-1} b_{n-2} \ldots$, where $z_k=0$
and $b_k$ is value of bit $b$ on step $k$. Let us consider set
of unitary gates for realization of a similar approach in quantum case.

Here $\ket{b}$ is qubit 
and continuous quantum variable may be represented via Hilbert space of 
complex functions $\psi(x)$ with real argument $x$ \cite{AV03,AV07,contvar}. 
Let us consider operator of conditional translation
\begin{equation}
\hat T = \ket{1}\bra{1} \otimes e^{-i \hat p /\hbar } + \ket{0}\bra{0} \otimes \hat 1.
\label{CT}
\end{equation}
This operator converts $\psi(x) \to \psi(x - 1)$, if qubit is in
state $\ket{1}$ and do nothing for $\ket{0}$.
Let us now introduce operator (projector) 
$\hat \Pi : \psi(x) \mapsto \Theta(x)\Theta(1-x) \psi(x)$,
where $\Theta(x)$ is the Heaviside step function
\[
 \Theta(x) = \begin{cases}
  1, & x \ge 0, \\
  0, & x < 0.
 \end{cases}
\]
It is possible to introduce conditional flip operator
\begin{equation}
\hat F = \bigl(\ket{1}\bra{0} + \ket{0}\bra{1}\bigr) \otimes \bigl(\hat 1 - \hat \Pi \bigr)
 + \hat 1 \otimes \hat\Pi.
\label{CFl}
\end{equation}
This operator flips state of qubit $\ket{b}$, if $\psi(x)$ {\em is zero 
on interval} $0 \le x \le 1$ and do nothing if the function is nonzero 
inside of this interval\footnote{It is also possible to use other operators,
{\em e.g.} sometimes it is more convenient to use an operator that flips state 
of qubit $\ket{b}$, only if $\psi(x)$ 
{\em is nonzero on interval} $1 \le x \le 2$.}.

It is possible to check, that for {\em function nonzero only inside this 
unit interval}, denoted further as $\psi_\sqcap(x)$, an operator 
$\hat T \hat F \hat T^*$ acts as (see Fig.~\ref{TFT})
\begin{equation}
 (\alpha \ket{0} + \beta \ket{1}) \otimes \psi_\sqcap(x) \mapsto
 \ket{0} \otimes (\alpha \psi_\sqcap(x) + \beta \psi_\sqcap(x-1)). 
\label{TFT}
\end{equation}

\begin{figure}[htb]
\begin{center}
\includegraphics[scale=0.32]{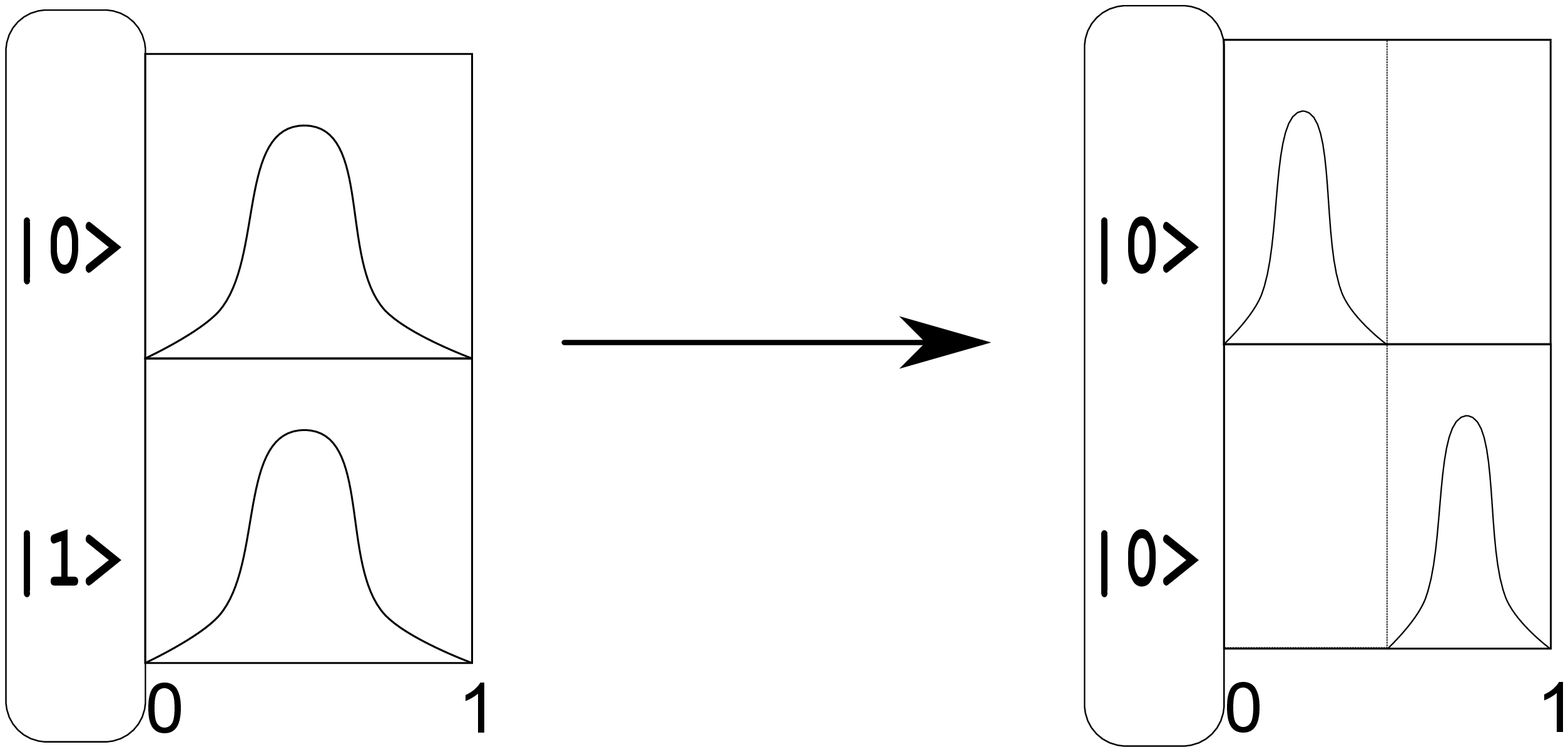} 
\end{center}
\caption{$\hat T \hat F \hat T^* \hat S$ for $0 \le x \le 1$ 
and basic states of qubit $\ket{0}$ or $\ket{1}$}\label{Fig:TFTS}
\end{figure}

Finally, it is possible to use squeezing operator \cite{contvar}
$\hat S : \psi(x) \mapsto \sqrt{2}\,\psi(2 x)$, 
{\em viz} $\hat S = \hat 1 \otimes e^{i (\hat x \hat p + \hat p \hat x) \ln 2 / 2\hbar}$.
After that both $\psi_\sqcap$ and ``shifted'' $\psi_\sqcap$
belong to initial unit interval (see Fig.~\ref{Fig:TFTS}).

\begin{figure}[htb]
\begin{center}
\includegraphics[scale=0.4]{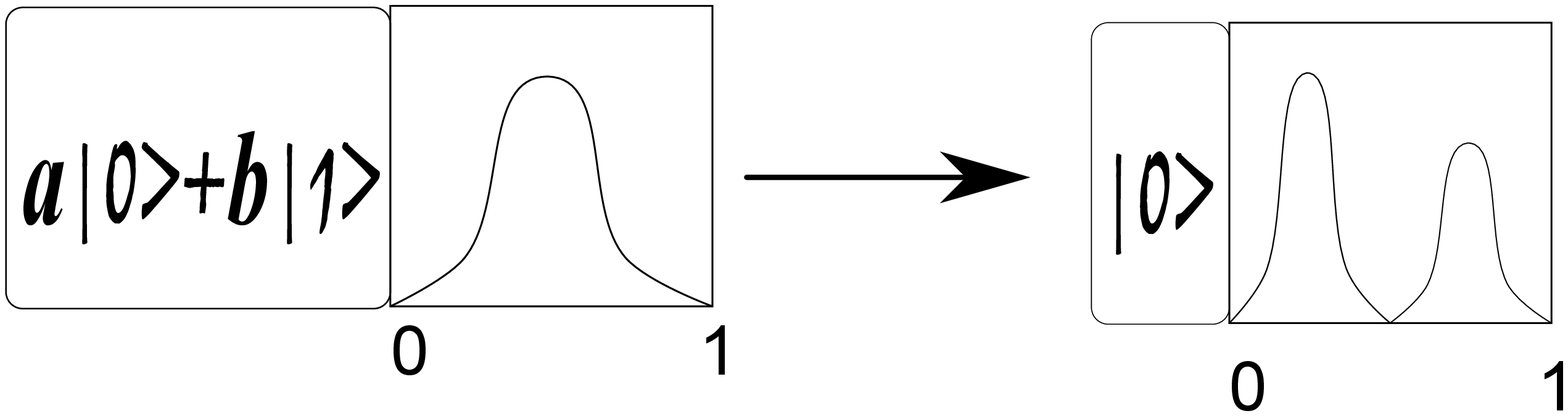} 
\end{center}
\caption{Operator $\hat E$ for $0 \le x \le 1$ 
and some qubit $a\ket{0}+b\ket{1}$}\label{Fig:Eab}
\end{figure}

Now, expression for ``erasure''
operator may be written as $\hat E = \hat T \hat F \hat T^* \hat S$ 
\begin{equation}
 \hat E: 
 (\alpha \ket{0} + \beta \ket{1}) \otimes \psi_\sqcap(x) 
 \mapsto \ket{0} \otimes \psi^{(\alpha,\beta)}_\sqcap(x), 
\label{ERAS}
\end{equation}
where  $\psi^{(\alpha,\beta)}_\sqcap(x) \equiv 
 \sqrt{2}\,\bigl(\alpha \psi_\sqcap(2x) + \beta \psi_\sqcap(2x-1)\bigr)$, 
see Fig.~\ref{Fig:Eab}.

\begin{figure}[htb]
\begin{center}
\includegraphics[scale=0.4]{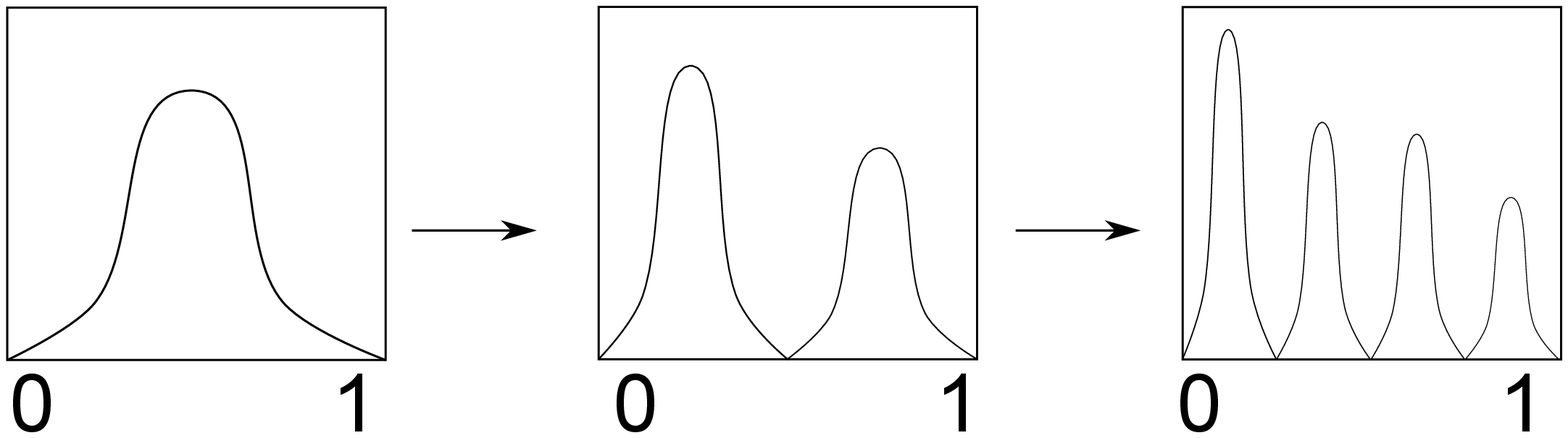} 
\end{center}
\caption{Transformation of a function $\psi_\sqcap(x)$
for consequent steps}\label{Fig:funs}
\end{figure}

If to start with continuous variable represented by an arbitrary 
function $\psi_\sqcap(x)$ {\em nonzero only on unit interval}, then 
after each step due to such transformation $\psi^{(\alpha,\beta)}_\sqcap(x)$ 
is another function {\em nonzero only on the same interval} and may be used for 
next step, see Fig.~\ref{Fig:funs}. 

Such a gate $\hat E$ can be used to attach continuous variables to all 
auxiliary qubits which need for ``cleaning'' after each step. Of course, 
such an operation may be used not only for programmable quantum processors, 
but here advantages are quite transparent due to necessity of numerous 
application of a standard operation.

\medskip

It should be mentioned, what for example with programmable quantum processor
such a ``cleaning'' is not always necessary. It is possible to use rather trivial
design with {\em read only memory} ({\sf ROM}) already discussed earlier \cite{AV01,AV03,AV07},
but in such a case it is not resembling usual universal computer.

\begin{figure}[htb]
\begin{center}
\includegraphics[scale=0.5]{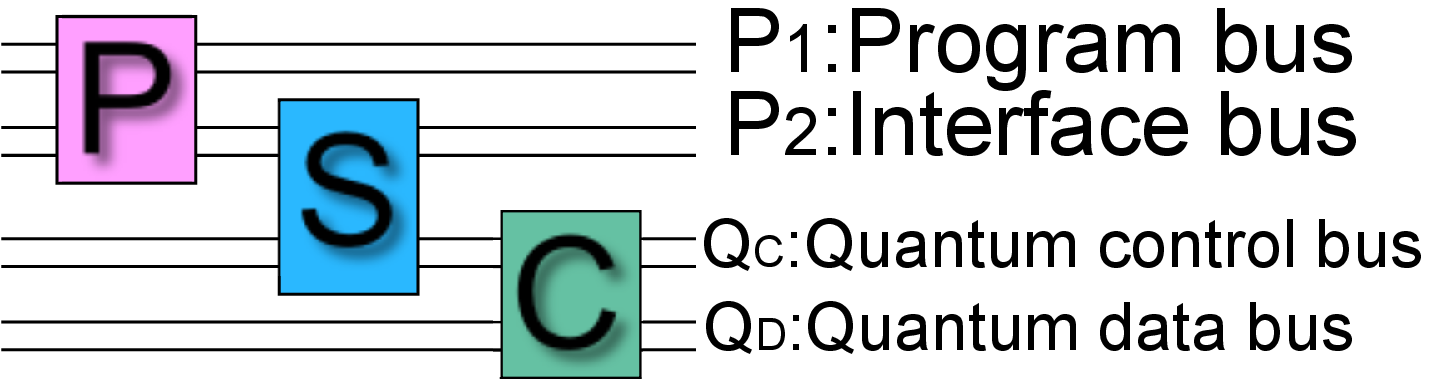} 
\end{center}
\caption{Programmable quantum processor}\label{Fig:pqp}
\end{figure}

Due to presented methods, design may be more similar 
with usual computing devices, Figure~\ref{Fig:pqp}. 
Here gate {\sf P} due to possibility to use any irreversible prototype may be an 
analogue of more or less traditional processors. Gates {\sf S} and {\sf C} are similar 
with ``{\sf Shift--Control}'' design discussed in \cite{AV03,AV07}, but instead of using 
cyclic {\sf ROM}, {\sf S} may obtain commands from interface bus of ``pseudo-classical 
processor'' {\sf P}.
 
In the \cite{AV03,AV07} were also discussed using continuous quantum variables
{\em in program register}. In such a case programmable quantum processor may
be {\em precisely} universal and for small number of qubits number of steps
may be limited by some reasonable number. 

And finally, it is possible to try develop reversible design of a processor from 
very beginning, to avoid necessity of consideration of irreversible operation,
but it is huge area of research and should be discussed elsewhere.




\end{document}